# Outsourcing in Global Software Development: Effects of Temporal Location and Methodologies

Mark Looi, University of Oxford Saïd Business School[1] *
Marc Szepan, University of Oxford Saïd Business School[2]

---

* Corresponding author
[1] Mark Looi, 3828 Cascadia Ave. S., Seattle, WA 98118, USA. Ph: +1 425 941 2378. Email: marklooi@gmail.com
[2] Marc Szepan, E2.17, W2.01, Saïd Business School, Park End Street, Oxford, OX1 1HP, UK. Ph: +44 7414 993030. Marc.Szepan@sbs.ox.ac.uk.

## List of Terminology and Abbreviations

Nearshore: Offshore outsourcing of software development to geographically Near or Far locations with Small temporal distances

Farshore: Offshore outsourcing of software development to geographically Far locations with Large temporal distances

GSE: Global Software Engineering

GSD: Global Software Development (often used synonymously with GSE)

## Abstract

Developing software globally using outsourced resources has become a common practice, with project teams often distributed in different time zones. In this study, we focus on customers that contract software development to vendors in temporally nearshore or far offshore locations. We conducted a survey to determine the effect of temporal distance on overall success, costs, project management effort, schedule, quality, communication problems, and other outcomes of interest to managers. In the survey of 80 customers and interviews with 6 of them, we also investigated the effect of software development methodology on the same outcomes. The results show that nearshore development is advantageous for overall success, quality, reduced PM effort, maintaining schedule, higher quality, and engendering fewer communication problems. Development methodology appears to only influence higher costs. We assess our findings in the context of prior GSE research and provide practical advice for customers of outsourced global software development, chief of which is to favor nearshore for communication-intensive or Agile projects.

## 1 Introduction

Global Software Engineering (GSE) or Development (GSD) has been studied for more than 20 years; in 2014, a taxonomy to describe the various combinations was described by (Šmite, et al., 2014). Using the taxonomy, we study *Offshore Outsourcing* of software development in locations with *Small* temporal distances (*nearshore*)[3] in comparison to locations with *Large* temporal distances (*farshore*) to a customer.

---

[3] The geographic distances could be Far (flying time greater than 2 hours) or Near (flying time less than 2 hours), but in this study, the temporal distance alone determines whether or not a location is *nearshore* (Šmite, et al., 2014).

Global software development has been used by managers for several decades to gain the strategic advantages of lower costs, access to more readily available skills, and local market expertise (Moe, et al., 2014). Outsourcing of software development is yet another management strategy prevalent since the 1970's (Šmite, et al., 2010). These two strategies are often combined (Šmite, et al., 2014) so that by outsourcing to vendors operating offshore, companies are able to quickly scale up development. Thus, software teams are often composed of both insourced and outsourced resources, either of which can operate onshore or offshore. Today, software engineering supply chains circle the globe to feed the rising demand for more software, which though increasing complex, must be produced faster and cheaper. Boeing's 737 Max crisis was laid at the feet of outsourcing software development to "$9-an-hour engineers" in India (Robison, 2019).

As the offshoring paradigm took hold and more locations for software development emerged, nearshore software development became an important variant. The rise of regional hubs and the desire to shorten supply chains (owing to trade wars and other geopolitical pressures) has underscored the merits of nearshore (Economist, 2019).

From the simple world where all software development by employees occurs under one roof, managers now could choose from a dozen combinations of Sourcing, Location, Legal entity, Geographic distance, and Temporal distance (Šmite, et al., 2014). Compounding the complexity for managers was the rapid shift to Agile development methodologies. A review of research shows that "there is no standard or recipe for successful GSE performance" (Šmite, et al., 2010). How should a manager approach such a bewildering choice?

This paper examines the circumstances under which offshore outsourcing to locations (geographically far or near) with small temporal differences (i.e., nearshore) should be preferred over far locations with large temporal distances (farshore). To address these questions, we review relevant current research in global software engineering as it applies to farshoring and nearshoring, then consider modern software practices such as Agile, and finally draw conclusions and recommendations based on our research. By design, we limit our study to offshore outsourcing patterns which are exclusively either nearshore or farshore (i.e., not a hybrid where a project might consist of both types of teams). The perspective of nearshore or farshore is exclusively from the point of view of the customer and the customer's location. Thus, a typical scenario that falls into our study would be a US-based firm that contracts with a third-party vendor based either in Argentina (nearshore) or India (farshore).

## 2 The Challenge of Outsourced Global Software Development

Firms often use a combination of insourced or outsourced teams, located farshore or nearshore to build their software (EY, 2013). These choices could have an impact on project outcomes and the research evidence so far is not encouraging. There are many examples of failures (Moe, et al., 2014), no clear indications of practices correlated with success (Šmite, et al., 2010), while recommendations for GSD success are unclear (Anh, et al., 2012).

Nearshore outsourcing of IT software development projects is often presumed to come with better team communication, though higher costs, compared to farshore (Boersen, et al., n.d.). The geographic distribution and temporal distance of development sites have a significant impact on quality (Cataldo & Nambiar, 2009). Our research investigates if the *amount* of temporal distance can affect outsourced project outcomes.

Strategies for software development methodologies fall into two broad camps: Agile and Structured. Agile uses an iterative process in which specification is kept high-level until work starts, at which time intensive, real-time coordination resolves details. Structured is a process where the software interfaces and features are well-defined, and teams can work independently on system components without the need for constant, real-time interaction. Research has suggested that development methodology has had little or no effect on outcomes for projects delivered by *globally distributed teams*, both insourced and outsourced (Estler, et al., 2014). Still, Agile in just 20 years has become the de facto standard methodology. Such a widespread acceptance of its efficacy suggests it is a determinant of success. Muddying the waters further, quality from methodology improvements seem to diminish with distributed development (Cataldo & Nambiar, 2009). Taking a narrower view, could there be a difference in outcomes among outsourced-only GSD projects depending on methodology?

In our study, *nearshore* locations are within *five* hours' time difference of the customer's location, though they could be still more distant geographically or culturally. We define farshore locations as farther away in time zone than five hours. We have chosen five hours despite the fact that some of the literature defines a nearshore *small* temporal distance as four or fewer hours (Šmite, et al., 2014). From our experience, the actual time difference can vary by time of year owing to observance of Daylight Savings Time in the northern hemisphere; so, to count all of South America as nearshore to the continental USA all year, it's useful to adjust the definition by one hour (i.e., Uruguay and Brazil are 4 hours ahead of Pacific Daylight Time, but are 5 hours ahead of Pacific Standard Time). Even with this minor adjustment, there is usually a meaningful overlap in the business day at each location such that extended co-working is possible; usually, there can be a 4 or more hours overlap with small adjustments in work shifts. With this adaptation, we utilize Šmite's taxonomy otherwise unchanged and narrow our study's context to temporal, not geographic distance (Šmite, et al., 2014). Thus, Argentina in geographically far from the US, but considered by US customers as nearshore, owing to temporal proximity.

Our research seeks to explore two questions: what is the effect of an outsourced temporal location (nearshore or farshore) on project outcomes in software development? And, what is the effect of development methodology on project outcomes in outsourced offshore (far or near) software development? Our research context—outsourced farshore or nearshore software development—avoids the broader categories of Information Technology Services (IT) or Business Process Outsourcing (BPO). We eschew considering *insourced* development whether offshore or not. By design, our inquiry sidesteps insourced and outsourced onsite software development, regardless of location or methodology. Thus, our focus is narrowed to outsourced software development in locations that are either temporally nearshore or farshore. The resulting context is still rich enough to gather data for our research. Though narrowed, these two management choices (location and methodology) confront managers regularly; we hope to provide them insights to help with these decisions.

Finally, our research focuses on the client or customer perspective exclusively (i.e., we eliminated survey data from vendors), something that most research does not sufficiently differentiate (Verner, et al., 2012).

This paper surveys relevant literature to review the research context, articulates the research questions, describes the investigation methodology, summarizes findings, discusses them, and makes recommendations for both customers engaging outsourced, offshore GSD and researchers.

# 3 Literature Review

In this section, we discuss the research context of outsourced offshore software development in both its variants, farshore and nearshore. We then pose our first question about the impact of outsourcing location on software development outcomes. Next, we introduce the concept of development methodologies most often used in software development and pose the second question—their impact on outcomes.

## 3.1 Outsourced Offshore Software Development

It is widely accepted that offshore software development, whether in-house or outsourced, is an arbitrage strategy aimed at lowering costs (Jiménez, et al., 2009) (Hanna & Daim, 2009), gaining access to skills, quality, and flexibility (Lacity, et al., 2016), or leveraging infrastructure without having to make a substantial investment (Khan, et al., 2010). Insourced offshore strategies tend to have better outcomes since internal teams emphasize minimizing cultural and trust barriers from the start, while overlooking them in offshore outsourcing (Prikladnicki & Audy, 2012).

Outsourced offshore development is not easy: it requires effective project management, relationship management, and cultural affinity (Khan, et al., 2012), while still encountering challenges such as cultural barriers, inadequacy of project management, weak IP protections, and technical capability (Khan, et al., 2011). National and organizational culture pose difficulties, too (Khan & Azeem, 2013). Top concerns identified by researchers are Communication, Group Awareness, Source Control, Knowledge Flow Management, Coordination, Collaboration, Project and Process Management, Process Support, Quality and Measurement, and Defects Detection (Jiménez, et al., 2009) (Jiménez & Piattini, 2008). Worse, the quality of outsourced offshore work is lower (Jabangwe, et al., 2015) and worsens the more distributed it is temporally and geographically (Cataldo & Nambiar, 2009).

As noted by Šmite, researchers have not always been careful about differentiating between offshore and farshore and other terms in common industry parlance are used inconsistently, making the relevant correlations hard to identify (Šmite, et al., 2014). We intend to avoid this ambiguity to concentrate on a narrow, but important context.

### 3.2  Outsourced Nearshore Software Development

Outsourced nearshore software development is the practice of locating the outsource delivery center temporally nearer the client (Carmel & Abbott, 2007). Hubs to serve the major advanced economies have emerged in the Americas, Central and Eastern Europe, and East Asia (Carmel & Abbott, 2006). Indeed, temporal proximity, lower transaction costs, and geographic nearness are the main reasons cited for nearshore (Oshri, et al., 2009).

Surprisingly, the advantages of nearshoring are less than clear. While it may offer opportunities for more successful collaboration especially for Agile methodologies, cost savings could be eroded (Šmite, et al., 2010). The literature around determinants of outsourcing success includes cultural distance, but not nearshore or farshore location (Lacity, et al., 2016). An open question remains: does outsourced nearshore development affect outcomes such as overall success, quality, schedule, etc., in comparison to farshore?

### 3.3  Software Development Methodologies

Software development methodologies range from simple coding standards to practices that span requirements gathering, design, coding, testing, deployment, maintenance and support (Pressman, 2005).

By the 1980's *structured software development methodologies* had become the mainstay for software development. Most firms followed some form of waterfall or other similar "structured" methodology (DeGrace & Stahl, 1990) (Boehm, 1988), which often feature well-defined processes, extensive specifications, detailed architecture and design, planning and project management (Pressman, 2005). Common criticisms of structured approaches were its long project durations, opacity of work-in-progress, and the misalignment of delivered features to actual need (Cohen, et al., 2004). Structured development continued to be refined by an industry of gurus who prescribed processes, best practices, standards and other ameliorations (McConnell, 1996) (Humphrey, 1989) (McCarthy, 1995). However, research suggested that non-agile methodologies caused development problems and that offshoring exacerbated them (Maxwell-Sinclair, 2016).

*Agile software development methodologies* were a response to these criticisms, for example SCRUM (Agile Alliance, 2018) (Schwaber, 2004). Agile approaches tend to emphasize frequent, informal, face-to-face communication among a small team of project members, including non-developers such as end users and product owners (Cohen, et al., 2004). They also avow incremental improvement centered around user scenarios, continuous delivery and a view that "working software is the primary measure of progress" (Beck, et al., 2001). Though articulated only in 2001, Agile quickly grew to 94% of all organizations reporting at least some use by 2017 (VersionOne, 2017).

Troubling signs emerged about the applicability of Agile in GSE. Jalili found that the Agile model had to be modified to accommodate GSE, but was unsure if those modifications still preserved the ethos of true Agile (Jalali & Wohlin, 2010). And, methodology alone could not sustain quality as development grew more distributed (Cataldo & Nambiar, 2009).

While the benefits and applicability of Agile, especially for co-located teams, were extensively researched and characterized (Estler, et al., 2014), in that same study of GSD, Estler concluded that there is "no significant difference between the outcome of projects following agile processes and structured processes … for globally distributed development." And, "the development process is not an independent variable" (Estler, et al., 2014). However, we suspect some aspects of the study may have contributed to these conclusions:

- Commingling of outsourced nearshore and farshore development projects
- Classification of a project into agile or structured by interpreting questionnaire answers
- Interviews of only Swiss companies

Given the Agile Manifesto's insistence on face-to-face meetings, the unimportance of location seems odd. What if we more narrowly confined our inquiry to purely outsourced software development projects where the customer is onshore, but the vendor is either farshore or nearshore? Could it then be that development methodology affects outcomes? Again, the literature around determinants of sourcing outcomes does not identify the role of development methodology (Lacity, et al., 2016).

In summary, our literature review shows that from the customer's point of view there are gaps in research about the effect of nearshore or farshore vendor locations on GSD outcomes and how methodology affects outsourced GSD outcomes.

### 3.4 Research Methods

Investigators in the literature cited have employed data collection methods ranging from semi-structured interviews (Weerakkody & Irani, 2009), surveys alone (Khan, et al., 2012), surveys with follow-up interviews (Estler, et al., 2014), content analysis (Carmel & Abbott, 2006) and systematic literature reviews (Khan, et al., 2011). As discussed in *Methodology,* our study utilizes surveys with follow-up interviews.

## 4    Research Questions

Offshoring in software development has been widely researched and analyzed. In a recent survey of software outsourcing literature, 27 determinants (or independent variables) were identified for outcomes (Lacity, et al., 2016). But, the role of nearshore versus farshore as one of the 27 determinants were rarely studied—just five occurrences in the review of 174 papers from 2010 to 2014 (Lacity, et al., 2016). Similarly, development methodology as a determinant of outcomes was not often cited. Our research attempts to fill these gaps by investigating both outsourcing location and development methodology's effects on outcomes. An adjunct to the second question looks at only Agile development to see if there are any further locational effects.

RQ1: In the outsourced IT Software Engineering industry, what is the effect of outsourced development team location (nearshore vs. farshore) on: 1) *Meeting frequency;* 2) *Communication problems;* 3) *Business day overlap;* 4) *Asynchronous communication;* 5) *Collaboration types;* 6) *Methodology;* 7) *Ways to define products;* 8) *Project management effort;* 9) *Success;* 10) *Cost;* 11) *Schedule;* 12) *Quality.*

RQ2: In the outsourced IT Software Engineering industry, what is the effect of software development methodology (Agile vs. Structured[4]) on: 1) *Location (nearshore vs. farshore);* 2) *Meeting frequency;* 3) *Communication problems;* 4) *Business day overlap;* 5) *Asynchronous communication;* 6) *Collaboration types;* 7) *Ways to define products;* 8) *Project management effort;* 9) *Success;* 10) *Cost;* 11) *Schedule;* 12) *Quality.*

RQ3: In the outsourced IT Software Engineering industry where the development methodology is Agile, what is the effect of outsourced development team location (nearshore vs. farshore) on: 1) *Meeting frequency;* 2) *Communication problems;* 3) *Business day overlap;* 4) *Asynchronous communication;* 5) *Collaboration types;* 6) *Ways to define products;* 7) *Project management effort;* 8) *Success;* 9) *Cost;* 10) *Schedule;* 11) *Quality.*

---

[4] The terms "Agile" and "Structured" are capitalized when they are referring to the methodologies.

For the purposes of our research and in the survey, nearshore is defined as five or fewer hours away from the respondent's central project locale. Any outsourcing relationship greater than five hours is therefore farshore. This departs slightly from the definition provided by (Šmite, et al., 2014), because as stated in section 2, to treat all of the Americas as one nearshore unit for customers in USA or Canada, it's necessary to allow up to a 5 hour time difference between Uruguay or Brazil and the Pacific Coast of North America (during the summer, this gap shrinks to 4 hours as the northern hemisphere observes Daylight Savings Time). Geographic and cultural proximity are not considered; nor are the use of multiple farshore and nearshore development locations in the same project.

The measures for our research questions reflect just a few of the "outcomes" (over 100) often studied in the literature; "outcomes" are usually characterized as dependent variables (Lacity, et al., 2016). However, one of our survey questions, "Ways to define products", is not often cited as a measure. We have reviewed these outcomes and used them in constructing the survey.

## 5  Methodology

Our research methodology consisted of a survey sent to subject-matter-experts with current or recent experience in the offshore outsourcing of GSD. After gathering and analyzing the responses, we interviewed some of them to discuss the quantitative findings to gain insights about motivations and underlying causes. Such a combination of quantitative and qualitative data gathering is a common methodology in social science research (Fowler, 2014).

## 5.1 Research Context

As described in the literature review, our overall research context is outsourcing; we narrow the context to IT and further still GSD. This context is still broad enough for us to gather data fairly easily.

Our research was conducted with senior managers or technical specialists in firms that engage in significant global software development using internal, outsourced, nearshore and farshore teams. More than 60% of the surveyed reported titles that were senior, typically requiring 10 or more years professional experience. These respondents can be expected to understand the differences between development strategies in industries as diverse as consumer services, banking, finance, airlines, technology, marketing, retail, consulting, government and others. Geographically, they hailed from the Americas, Europe, Middle East and Asia. Our study sought input from a diversity of industries, geographies and professional roles because of the ubiquity of software in modern business and our intent to use a relativistic frame of reference for nearshore and farshore inquiry. That is, a nearshore team for a customer in Paris could well be farshore to a manager in Seattle; while the location of the team is invariant, the outcomes could differ depending, we postulated, on the location of the customer. With this approach, it is intended that the data reflect more than a particular local or national perspective. Thus, our research context should elicit results that are potentially applicable to different industries, geographies, and are somewhat culturally agnostic.

## 5.2 Hypotheses

We construct null hypotheses for each of the 3 research questions; a null hypothesis posits the lack of a correlation between outcome and temporal location or methodology. That is, the null hypothesis for our RQ1 is: $H_{0,\ location}^{i}$, $i \in \{1, \ldots 12\}$, for GSD nearshore or farshore projects, there is no difference in:

1. *Meeting frequency*
2. *Communication problems*
3. *Business day overlap*

4. *Asynchronous communication*
5. *Collaboration*
6. *Methodology*
7. *Ways to define products*
8. *Project management effort*
9. *Success*
10. *Cost*
11. *Schedule*
12. *Quality*

So, for example, the formal expression of $H_{0,\ location}^{1}$ is: There is no difference in *Meeting frequency* for GSD nearshore or farshore projects.

RQ2 leads to another set of null hypotheses; $H_{0,\ methodology}^{i}, i \in \{1, \ldots 12\}$, for GSD projects developed using Agile vs. Structured methods, there is no difference in:

1. *Location*
2. *Meeting frequency*
3. *Communication problems*
4. *Business day overlap*
5. *Asynchronous communication*
6. *Collaboration*
7. *Ways to define products*
8. *Project management effort*
9. *Success*
10. *Cost*
11. *Schedule*
12. *Quality*

And, a third set of null hypotheses, $H_{0,\ Agile\&location}^{i}, i \in \{1, \ldots 11\}$, for Agile-only GSD projects done nearshore or farshore, there is no difference in:

1. *Meeting frequency*

2. *Communication problems*
3. *Business day overlap*
4. *Asynchronous communication*
5. *Collaboration*
6. *Ways to define products*
7. *Project management effort*
8. *Success*
9. *Cost*
10. *Schedule*
11. *Quality*

Our study then collects survey data that may falsify some or all of these null hypotheses, thereby asserting the alternative hypothesis: that location or methodology has an effect on outcomes in outsourced software development.

## 5.3 Survey Questionnaire

We asked survey respondents to consider outsourced software development projects that they are familiar with and that was developed either nearshore or farshore. We then asked them about the following management choices, processes or results:

- Software development methodologies
- Real-time meetings (e.g., video, audio or in-person)
- Collaboration type (e.g., asynchronous or synchronous)
- Asynchronous meeting adequacy
- Arbitrage outsourcing strategies
- Outcomes of project in question

The questions correspond to the null hypotheses and are reproduced in *Appendix A: Survey Questions.* They were selected to draw out responses to outcomes managers are most interested in. The survey can be completed in less than 5 minutes with an estimated 60% or greater completion rate. We followed common survey best practices (Looi, 2017) to encourage participation while forcing respondents to make clear, pre-defined choices where required by our study (e.g., location). Some questions require respondents to rate an experience, like success, on a 5-point Likert scale. To improve validity, we piloted and tested the survey on three subjects familiar with survey design and software engineering before launching it to the target audience.

## 5.4 Selection of Participants

We chose participants who are qualified managers or software professionals functioning in a customer capacity (i.e., buyers of outsourcing software development services). Some participants had 30 years' experience in the field; based on titles, they were senior staff. Using postings on social media (e.g., LinkedIn, Twitter), emails to professional networks, industry interest distribution lists, and personal contacts, we obtained 143 survey respondents from Asia, the Americas, and Europe. Of these, 63 self-disqualified because they could not abide the terms of the survey (e.g., they were not customers) or determined they didn't have the knowledge to complete it. The survey was quite strict about ensuring that the respondents were in fact on the customer side of outsourcing and that they used either farshore or nearshore vendors. For example, a manager who only had experience with insourced farshore GSD would have their responses disqualified.

*Figure 1* depicts the roles of qualified survey respondents; *Figure 2* shows the firm size.

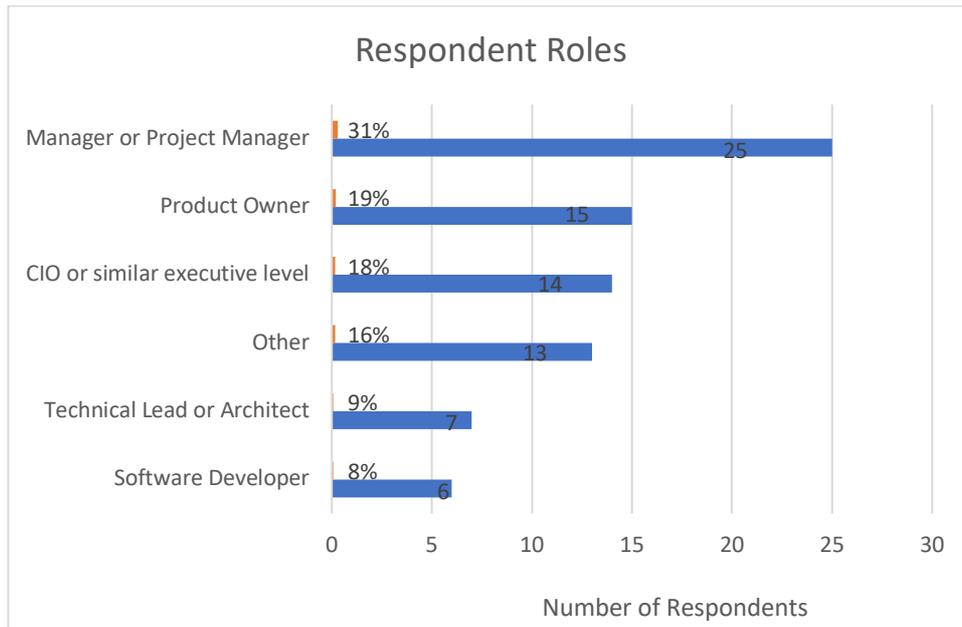

*Figure 1: Respondent Roles*

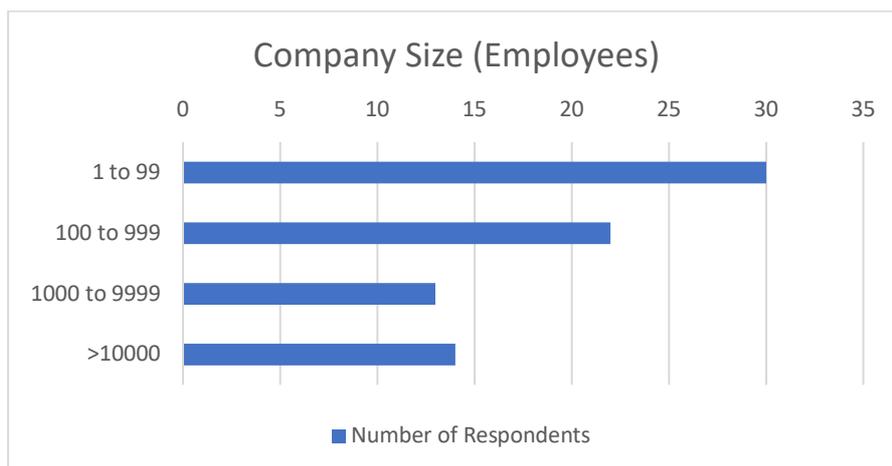

*Figure 2: Respondent Company Size*

## 5.5 Data Collection

Our online survey was open for two weeks during which time at least 2500 people were exposed to the survey request (there could have been many more because of the use of publically-visible hashtags). The survey could be completed using a Web or mobile browser or the SurveyMonkey app. The targeted cohort was reminded at least once. Data was collected using only SurveyMonkey, where the raw data is archived. The respondents were informed that the survey was in support of academic research at the University of Oxford.

In a follow-on step, we interviewed 6 of the survey-takers in person, videoconference, or teleconference. The outline of questions is shown in *Appendix B: Qualitative Interview Questions.* Interviewees had the following roles: Software Engineer, Senior Director, Senior VP, Managing Director, CEO, and Chief Data Officer. Their firms were located in Doha, Qatar; Toronto, Canada; Los Angeles, USA; Cheltenham, UK; Paris, France and Seattle, WA; London, UK. Interviewees' roles and locations were chosen to foster heterogeneity.

## 5.6 Data Analysis

We performed both correlation and hypotheses tests. The purpose of correlation tests is to see which variables might be predictive of each other and to check that they are internally consistent (Harmon, 2011) (Mood, et al., 1974). We can then concentrate on those hypotheses that are likely to yield interesting conclusions. Since we cannot assume a normal distribution of data, we use Kendall's τ correlation test (Padgett, 2011) (Wessa, 2018).

### 5.6.1 Hypotheses Tests

For null hypotheses, $H_0^i : P(x > y) = P(x < y)$, where $i \in \{1, \ldots N\}$ in the list of hypotheses and $(x, y) \in \{$ (Agile, Structured), (Nearshore, Farshore), ((Nearshore, Farshore) | Agile) $\}$, we use the $\chi^2$ Test to reject any hypothesis where $p < \alpha$, the significance level. This test is used because our hypothesis data consist of dependent and independent variables with categorical data, (UCLA Institute for Digital Research and Education, 2017). For our tests we take $\alpha = 0.05$, a value commonly used in social sciences (Fisher, 1934). For any rejected hypotheses, we accept their alternatives, $H_1^i : P(x > y) \neq P(x < y)$.

For each rejected hypothesis, to determine if differences between means are statistically significant, we apply a single tail, heteroscedastic T-Test (Hays & Winkler, 1970). However, the T-Test assumes normality; to ascertain that, we use the Shapiro-Wilk Test (Dittami, 2018) to examine if the response frequency counts are normally distributed; where there is a large enough sample size (above 30), we depend on the Central Limit Theorem to assume normality (Hays & Winkler, 1970).

The analysis is then charted and summarized into tables. Microsoft Excel was used to perform most of the calculations, except where noted.

## 6   Findings

We analyzed the responses statistically, confirming or rejecting hypotheses. A further set of questions was constructed to elicit motivations or reasons for the survey results. We then interviewed some of the respondents to obtain clarification and detail around their experience with nearshore or farshore outsourcing and with development methodologies in the context of outsourcing. Our analysis showed that nearshore location has an effect on *overall success, costs, Project Management (PM) effort, schedule, quality,* and *communication problems.* Agile software development seems to increase *costs.* In this section, we summarize the data, describe the hypotheses tests results, present our quantitative conclusions, and describe the insights gained from interviews.

### 6.1   Data

We present the survey data in graphs, showing the number of responses by independent variables, nearshore and farshore, for each question in the survey in *Figure 3.* A perusal of the data suggests some marked differences between nearshore and farshore in measures such as *collaboration type*, but there are indications of a correlation between nearshore and *success.*

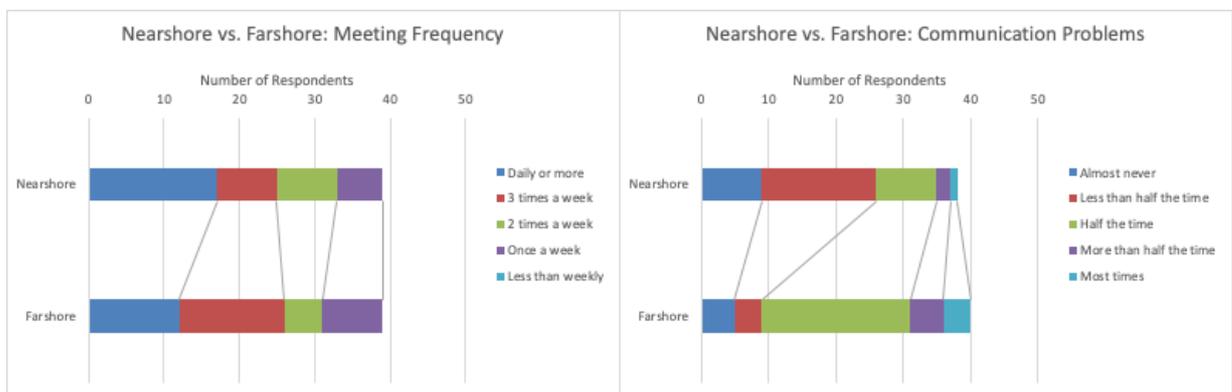

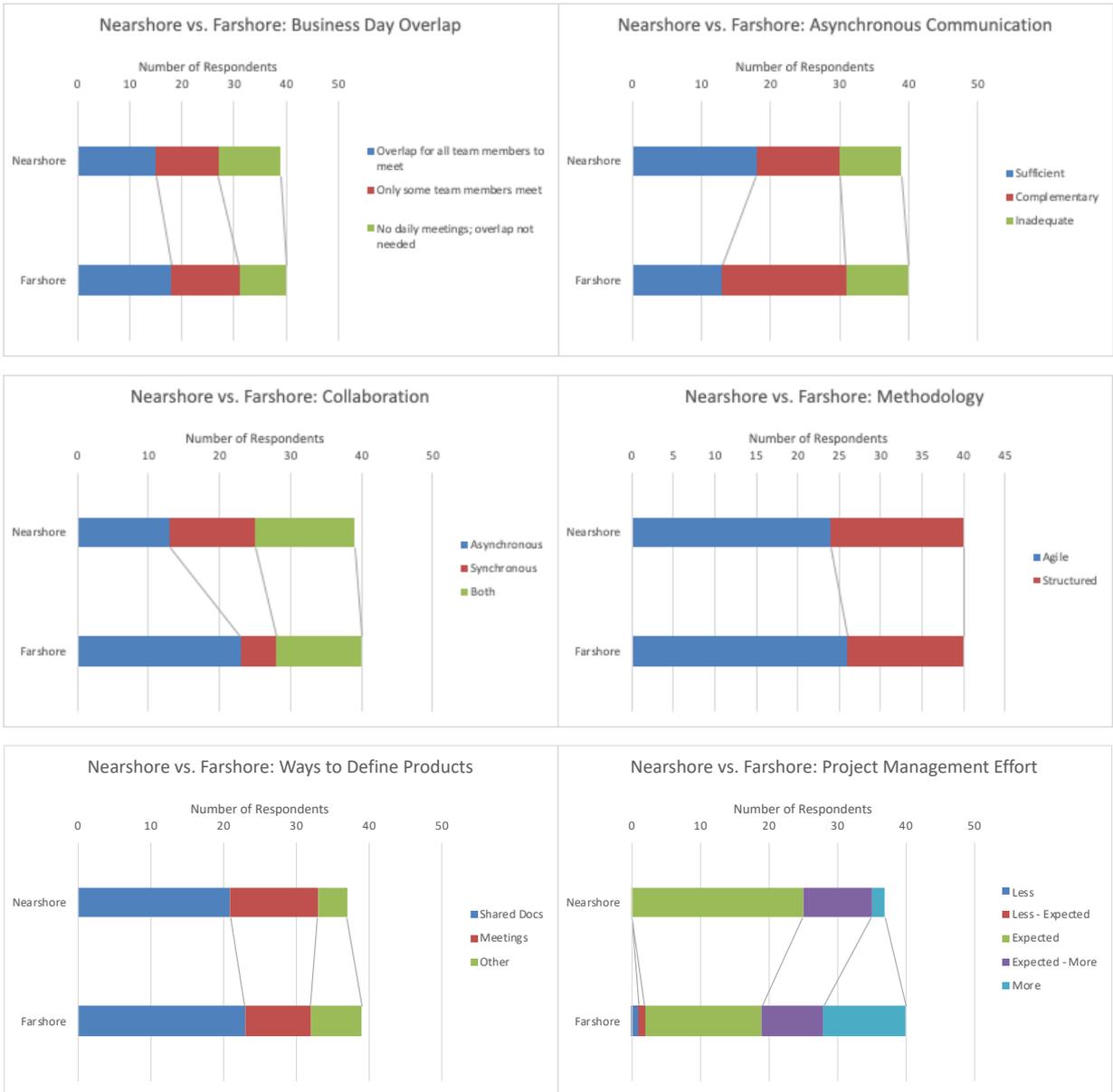

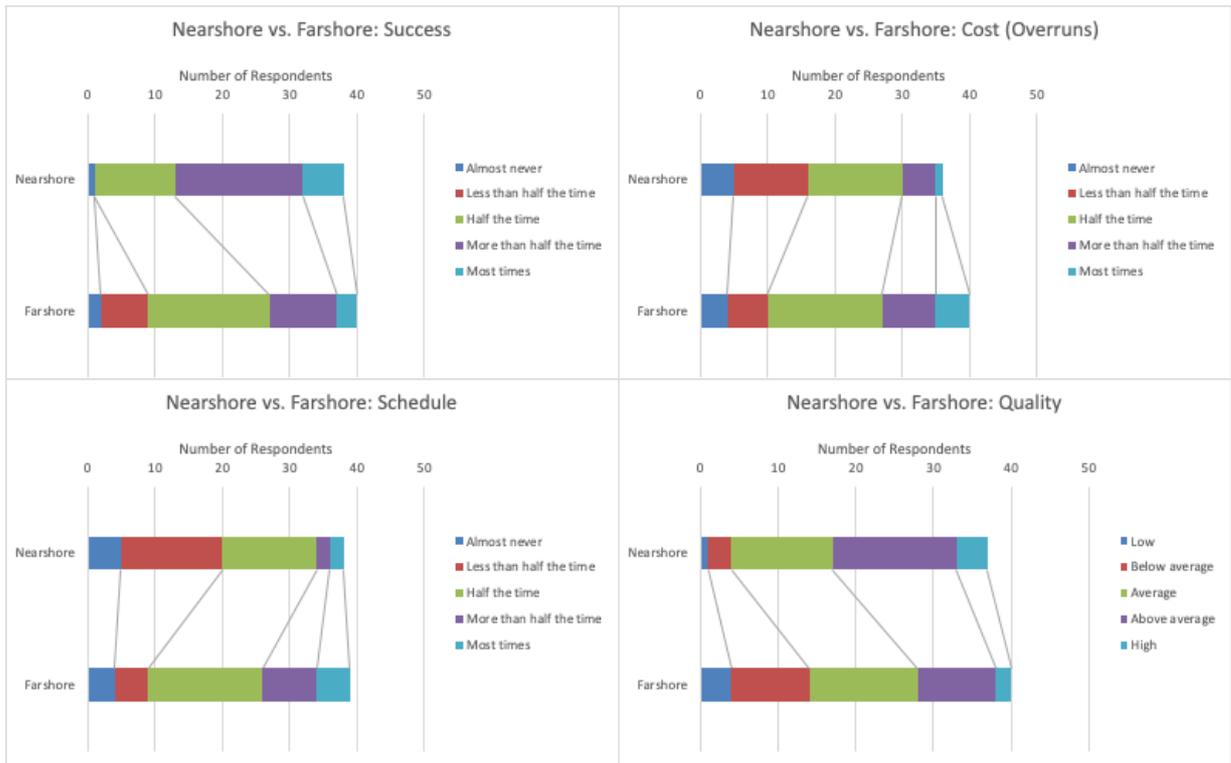

*Figure 3: Data for Location Hypotheses*

*Figure 4* show shows the same survey data graphed to show the relationship of methodology by various outcome measurements. Reflecting its ascendency, there were substantially more respondents reporting the use of Agile methodologies. Given this asymmetry, it is not immediately obvious that there is much of a difference between the Agile and Structured across the various measures.

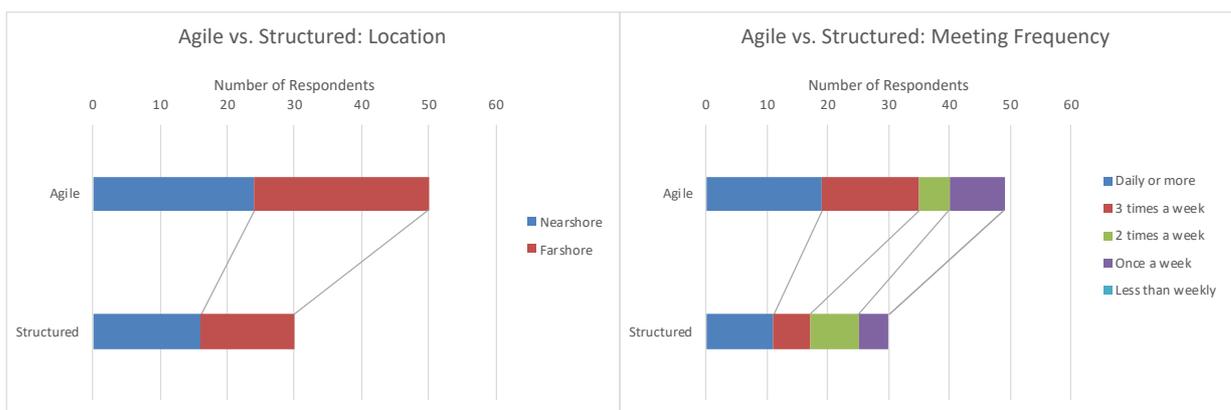

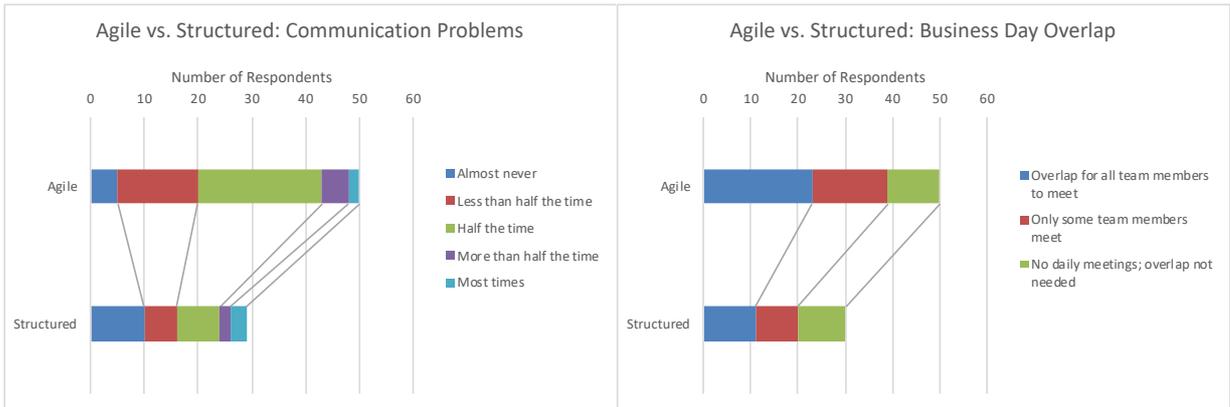
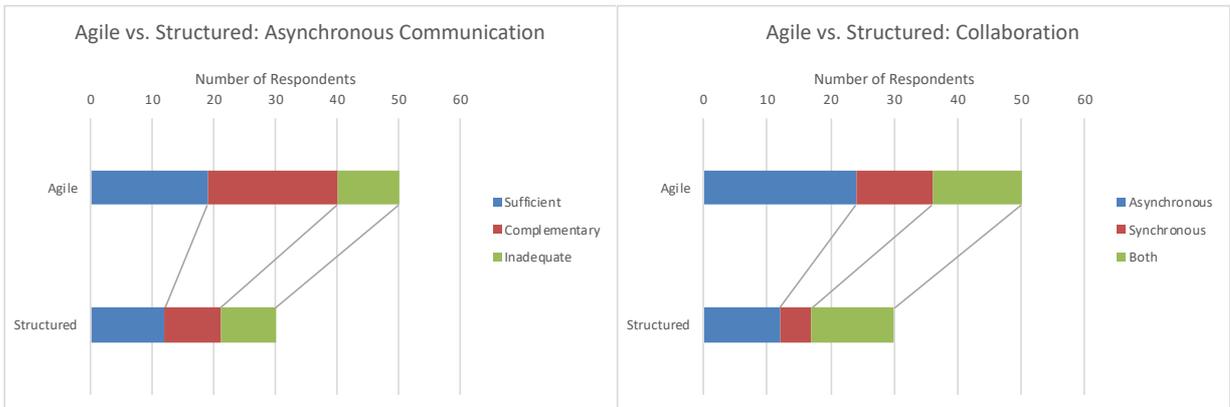
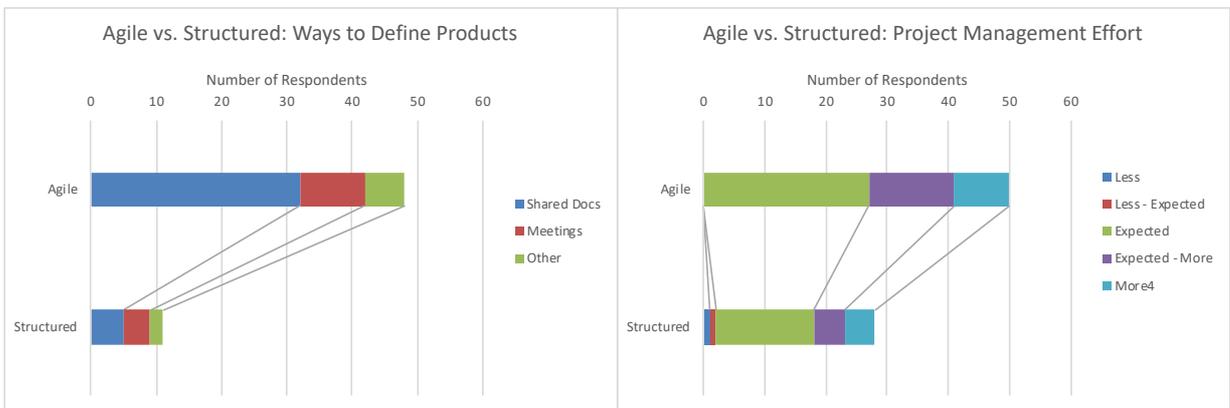

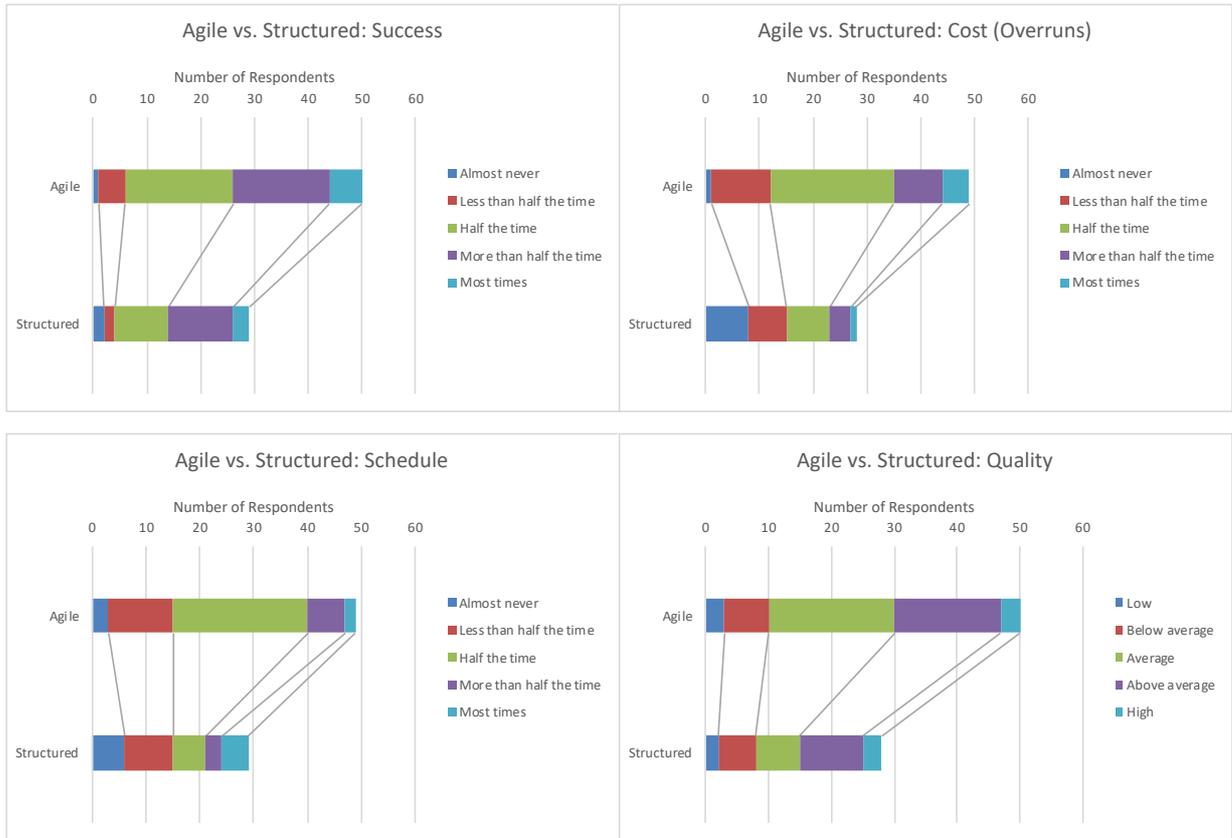

*Figure 4: Data for Methodology Hypotheses*

To see relationships more clearly, we must proceed to correlations and hypothesis testing.

### 6.1.1 Correlations

We begin by determining the correlation between some of the variables we have observed. Most of the variable pairs show no substantial correlation. *Table 1* shows the variable pairs that had correlation where $|\tau| \geq 0.3$ (Wessa, 2018).

| Variable Pairs | | $\tau$ | $p$ |
|---|---|---|---|
| Success | Location: Farshore | -0.336 | < 0.002 |
| Success | Communication Problems | -0.335 | < 0.001 |
| Success | Schedule Problems | -0.328 | < 0.001 |

*Table 1: Kendall Correlation of Variable Pairs, $|\tau| \geq 0.3$*

Of all the outcome-related measures, *success* summarizes other measures. Indeed, it correlates negatively with *communication problems* and *schedule*. These correlations are not surprising and serve as partial verification of the consistency of respondents' answers. *Farshore* location correlates negatively with *success*; it suggests that location could have an effect on *success* (and its correlates). To examine these further, we must test the hypotheses.

## 6.2 Analysis of Hypotheses

For null hypotheses, $H_0^i : P\,(\,x > y\,) = P\,(\,x < y\,)$, where $i \in \{\,1, \ldots N\,\}$ in the hypotheses $(\,x, y\,) \in \{$ (Agile, Structured), (Nearshore, Farshore), ((Nearshore, Farshore) | Agile) $\}$, we provide the mean, variance, T-Test, $\chi^2$ Test, and histogram. We also show which of them can be rejected because $p < 0.05$. For any rejected hypotheses, we accept their alternatives, $H_1^i : P\,(\,x > y\,) \neq P\,(\,x < y\,)$.

*Table 2*[5] summarizes the Location = (Nearshore, Farshore) results.

---

[5] Some cells in this and other tables are empty because the calculations make no sense—such as the mean of a categorical variable.

| i | Null Hypothesis | Location | Count | μ | Histogram | σ | T-Test | | p | Reject |
|---|---|---|---|---|---|---|---|---|---|---|
| 1 | Meeting frequency (5+, 3, 2 1, <1 times a week) | Nearshore | 40 | | | | | 3.4286 | 0.18009 | No |
| | | Farshore | 40 | | | | | | | |
| 2 | Communication problems (1:Never - 5:Most Times) | Nearshore | 39 | 2.154 | | 0.97645 | 0.00047 | 25.4993 | 2.90E-06 | Yes |
| | | Farshore | 40 | 2.975 | | 1.07387 | | | | |
| 3 | Business day overlap (Sufficient, Adequate, NA) | Nearshore | 40 | | | | | 0.5862 | 0.74594 | No |
| | | Farshore | 40 | | | | | | | |
| 4 | Asynchronous communication (Sufficient, Complementary, Not) | Nearshore | 40 | | | | | 2.0591 | 0.35717 | No |
| | | Farshore | 40 | | | | | | | |
| 5 | Collaboration types: Async, Sync, Both | Nearshore | 40 | | | | | 5.9935 | 0.04995 | Yes |
| | | Farshore | 40 | | | | | | | |
| 6 | Agile or Structured | Nearshore | 40 | | | | | 0.2133 | 0.64417 | No |
| | | Farshore | 40 | | | | | | | |
| 7 | Ways to define products: Docs, Meetings, Other | Nearshore | 37 | | | | | 1.2859 | 0.52573 | No |
| | | Farshore | 39 | | | | | | | |
| 8 | Project management effort (1:Less - 5:More) | Nearshore | 38 | 3.368 | | 0.59401 | 0.02540 | 6.2911 | 0.04304 | Yes |
| | | Farshore | 40 | 3.750 | | 1.00639 | | | | |
| 9 | Success (1:Never - 5:Most Times) | Nearshore | 39 | 3.769 | | 0.81983 | 0.00117 | 11.9226 | 0.00258 | Yes |
| | | Farshore | 40 | 3.125 | | 0.96576 | | | | |
| 10 | Higher Cost (1:Never - 5:Most Times) | Nearshore | 37 | 2.595 | | 0.96937 | 0.02409 | 10.1113 | 0.00637 | Yes |
| | | Farshore | 40 | 3.100 | | 1.12774 | | | | |
| 11 | Schedule (1:Never - 5:Most Times) | Nearshore | 39 | 2.487 | | 0.97916 | 0.00546 | 0.0072 | 0.01358 | Yes |
| | | Farshore | 39 | 3.128 | | 1.12810 | | | | |
| 12 | Quality (1:Low - 5:High) | Nearshore | 38 | 3.526 | | 0.90045 | 0.00378 | 0.0183 | 0.02399 | Yes |
| | | Farshore | 40 | 2.900 | | 1.05733 | | | | |

*Table 2: Location Hypotheses Summary*

The null hypotheses we reject are $H_0^i : i \in \{2, 5, 8, 9, 10, 11, 12\}$; or, *communication problems, collaboration types, project management effort, success, cost, schedule,* and *quality.* The Shapiro-Wilk Test shows that the response frequency counts are normally distributed for the *success* and *quality* hypotheses; hypotheses 2, 5, 8, 10, 11 have a large enough sample size to assume normality. The single tail, heteroscedastic T-Test for each of these rejected hypotheses shows that the differences between $\mu_{Nearshore}$ and $\mu_{Farshore}$ are statistically significant (Hays & Winkler, 1970). The means and their variances are plotted in *Appendix C: Means and Standard Deviations.*

*Table 3* shows the Methodology = (Agile, Structured) results.

| i | Hypothesis | Methodology | Count | μ | Histogram | σ | T-Test | | p | Reject |
|---|---|---|---|---|---|---|---|---|---|---|
| 1 | Nearshore or Farshore | Agile | 50 | | | | | 0.5020 | 0.64417 | No |
| | | Structured | 30 | | | | | | | |
| 2 | Meeting frequency (5+, 3, 2 1, <1 times a week) | Agile | 50 | | | | | 2.2857 | 0.31891 | No |
| | | Structured | 30 | | | | | | | |
| 3 | Communication problems (1:Never - 5:Most Times) | Agile | 50 | 2.680 | | 0.93547 | 0.14316 | 1.7035 | 0.19184 | No |
| | | Structured | 29 | 2.379 | | 1.32055 | | | | |
| 4 | Business day overlap (Sufficient, Adequate, NA) | Agile | 50 | | | | | 0.9041 | 0.63632 | No |
| | | Structured | 23 | | | | | | | |
| 5 | Asynchronous communication (Sufficient, Complementary, Not) | Agile | 50 | | | | | 1.5288 | 0.46561 | No |
| | | Structured | 30 | | | | | | | |
| 6 | Collaboration types: Async, Sync, Both | Agile | 50 | | | | | 2.0473 | 0.35927 | No |
| | | Structured | 30 | | | | | | | |
| 7 | Ways to define products: Docs, Meetings, Other | Agile | 48 | | | | | 1.7648 | 0.41379 | No |
| | | Structured | 11 | | | | | | | |
| 8 | Project management effort (1:Less - 5:More) | Agile | 50 | 3.640 | | 0.77618 | 0.16182 | 1.6990 | 0.37776 | No |
| | | Structured | 28 | 3.429 | | 0.95950 | | | | |
| 9 | Success (1:Never - 5:Most Times) | Agile | 50 | 3.460 | | 0.90824 | 0.16182 | 0.0542 | 0.81729 | No |
| | | Structured | 29 | 3.414 | | 1.01831 | | | | |
| 10 | Higher Cost (1:Never - 5:Most Times) | Agile | 49 | 3.122 | | 0.94940 | 0.00349 | 32.7158 | 7.87E-08 | Yes |
| | | Structured | 28 | 2.393 | | 1.16553 | | | | |
| 11 | Schedule (1:Never - 5:Most Times) | Agile | 49 | 2.857 | | 0.88976 | 0.32276 | 1.3853 | 0.34055 | No |
| | | Structured | 29 | 2.724 | | 1.38607 | | | | |
| 12 | Quality (1:Low - 5:High) | Agile | 50 | 3.200 | | 0.96890 | 0.47772 | 0.6637 | 0.58145 | No |
| | | Structured | 28 | 3.214 | | 1.13389 | | | | |

*Table 3: Methodology Hypotheses Summary*

The only null hypothesis that can be safely rejected is *cost.* The single tail, heteroscedastic T-Test for the rejected hypothesis shows that the difference between $\mu_{Agile}$ and $\mu_{Structured}$ is statistically significant, indicating Agile costs are higher. For this hypothesis only, the distribution of the response frequency counts is also normally distributed. The means and their variances are plotted in *Appendix C: Means and Standard Deviations.*

*Table 4* summarizes results where the Location = (Nearshore, Farshore), given Agile is the assumed methodology.

| i | Null Hypothesis | Location | Count | μ | Histogram | σ | T-Test | | p | Reject |
|---|---|---|---|---|---|---|---|---|---|---|
| 1 | Meeting frequency (5+, 3, 2 1, <1 times a week) | Nearshore | 23 | | | | | 0.6734 | 0.41187 | No |
| | | Farshore | 25 | | | | | | | |
| 2 | Communication problems (1:Never - 5:Most Times) | Nearshore | 23 | 2.435 | | 0.94514 | 0.02232 | 18.4484 | 0.00010 | Yes |
| | | Farshore | 26 | 2.962 | | 0.82369 | | | | |
| 3 | Business day overlap (Sufficient, Adequate, NA) | Nearshore | 23 | | | | | 1.0703 | 0.58557 | No |
| | | Farshore | 26 | | | | | | | |
| 4 | Asynchronous communication (Sufficient, Complementary, Not) | Nearshore | 23 | | | | | 0.4102 | 0.81457 | No |
| | | Farshore | 26 | | | | | | | |
| 5 | Collaboration types: Async, Sync, Both | Nearshore | 23 | | | | | 3.9079 | 0.14171 | No |
| | | Farshore | 26 | | | | | | | |
| 6 | Ways to define products: Docs, Meetings, Other | Nearshore | 22 | | | | | 2.9193 | 0.23231 | No |
| | | Farshore | 25 | | | | | | | |
| 7 | Project management effort (1:Less - 5:More) | Nearshore | 23 | 3.348 | | 0.57277 | 0.00362 | 4.7405 | 0.02946 | Yes |
| | | Farshore | 26 | 3.923 | | 0.84489 | | | | |
| 8 | Success (1:Never - 5:Most Times) | Nearshore | 23 | 3.870 | | 0.75705 | 0.00076 | 8.7797 | 0.01240 | Yes |
| | | Farshore | 26 | 3.077 | | 0.89098 | | | | |
| 9 | Higher Cost (1:Never - 5:Most Times) | Nearshore | 22 | 2.773 | | 0.92231 | 0.00545 | 7.8857 | 0.01939 | Yes |
| | | Farshore | 26 | 3.462 | | 0.85934 | | | | |
| 10 | Schedule (1:Never - 5:Most Times) | Nearshore | 23 | 2.435 | | 0.78775 | 0.00028 | 10.3104 | 0.00577 | Yes |
| | | Farshore | 25 | 3.280 | | 0.79162 | | | | |
| 11 | Quality (1:Low - 5:High) | Nearshore | 23 | 3.652 | | 0.71406 | 2.76923 | 11.1634 | 0.00377 | Yes |
| | | Farshore | 26 | 2.769 | | 0.99228 | | | | |

*Table 4: Agile and Location Hypotheses Summary*

As in the earlier set of null hypotheses involving location, when we assume Agile methodology, the same hypotheses are rejected—there is no difference in: *success, cost, project management effort, schedule, quality,* and *communication problems*. In addition, *success, cost, schedule,* and *quality* are normally distributed and a single tail, heteroscedastic T-Test shows that their means have a statistically significant difference in all cases except *quality*.

*Table 5* compares the means of the total dataset and Agile-only, for normally distributed variables. The purpose of the comparison is to get a directional indication if an Agile strategy can affect outcomes—it is not to compute quantitative differences in methodologies.

| i | Null Hypothesis | Location | μ (All) | μ (Agile) | μ (All) - μ (Agile) | Agile Higher |
|---|---|---|---|---|---|---|
| 1 | Success (1:Never - 5:Most Times) | Nearshore | 3.769 | 3.870 | -0.1003 | Yes |
|   |   | Farshore | 3.125 | 3.077 | 0.0481 | No |
| 2 | Higher Cost (1:Never - 5:Most Times) | Nearshore | 2.595 | 2.773 | -0.1781 | Yes |
|   |   | Farshore | 3.100 | 3.462 | -0.3615 | Yes |
| 3 | Project management effort (1:Less - 5:More) | Nearshore | 3.368 | 3.348 | 0.0206 | No |
|   |   | Farshore | 3.750 | 3.923 | -0.1731 | Yes |
| 4 | Schedule (1:Never - 5:Most Times) | Nearshore | 2.487 | 2.435 | 0.0524 | No |
|   |   | Farshore | 3.128 | 3.280 | -0.1518 | Yes |
| 5 | Quality (1:Low - 5:High) | Nearshore | 3.526 | 3.652 | -0.1259 | Yes |
|   |   | Farshore | 2.900 | 2.769 | 0.1308 | No |
| 6 | Communication problems (1:Never - 5:Most Times) | Nearshore | 2.154 | 2.435 | -0.2809 | Yes |
|   |   | Farshore | 2.975 | 2.962 | 0.0135 | No |

*Table 5: Comparison of Means*

The comparisons show that for nearshore development, when compared to the population as a whole, Agile methodology has better *success*, lower *PM effort,* and higher *quality*; it also had higher *costs* and *communication problems* (see final column, "Agile Higher"). Agile farshore development had lower *success*, greater *PM effort,* more *schedule* problems, and lower *quality*, when compared to the combined population of Agile and Structured development. Likelihood of *cost* increases was higher in both near and far, but Agile farshore software development showed the largest increase. In summary, when isolating Agile-only software development, the same characteristics are evident as in the overall dataset, just more pronounced.

In conclusion, the analysis shows that:

- A nearshore location for outsourcing of software development results in statistically significant desirable differences in overall *success,* lower *cost*, lower *Project Management effort,* maintaining *schedule,* higher *quality,* and fewer *communication problems.*

- Agile software development results in statistically significant increased *cost.* It has no statistically significant effect on any other variable we measured.

- Holding Agile methodology constant does not alter *nearshore* location effects, but it was associated with more negative outcomes already found in farshore software development while emphasizing the positive ones. As expected, Agile was correlated with greater *cost* risk generally but had a stronger effect on *farshore*.

### 6.3 Interviews

To seek underlying motivations and insights about our quantitative findings, we interviewed 6 survey respondents. The questions asked are shown in *9.* As noted in *Data Collection,* their roles and locations were diverse. Responses were inductively coded and summarized in *Table 6,* with a count of the codes mentioned by the interviewees during the one-on-one interview.

| Question | Responses [count of mentions] |
|---|---|
| Outsourcing nearshore motivations | Cost [4]. Cultural affinity [2]. Talent [4]. Flexibility & Availability [2]. Communication [1]. Complex systems better done nearshore [1]. Local market knowledge [2]. |
| Outsourcing farshore motivations | Flexibility & Availability [1]. Cost [2]. Talent [3]. Resources [1] Schedule [1]. Round-the-clock work [1]. Local market knowledge [1]. |
| Reasons for using location | Nearshore: Support [2]. Cost [1]. Talent [1]. <br> Farshore: Software development [1]. Talent [2]. |
| Accommodations for outsourcing | Farshore: limited synchronous communication [1]. Project leader nearshore to assist farshore team [1]. Business day time-shifting [3]. Invest in cultural alignment, often neglected with vendors [1]. Collaboration tools [5]. SLAs [1]. |
| Observed nearshore success | Yes [5]. |
| Reasons for nearshore success | Cost [1]. Efficiency [3]. Quality [1]. Responsiveness [1]. Better adherence to schedule [1]. Cultural affinity [1]. Better communication [4]. <br> Farshore more transactional [1]; more difficult [1]. |
| Reasons for firm's methodology | Agile: Flexibility [4]. Responsiveness [5]. Vendor accountability [1]. Transparency [1]. Honest conversations, team cohesion/dynamics [2]. Better quality [1]. Fits startup paradigm [1]. Develop internal capabilities [1]. <br> Structured: good for outsourcing [1] |
| Reasons for higher Agile costs | Requirements changes [3]. Work aborted [2]. Cost of quality underestimated at outset [1]. Communications intensive [3]. Product-market fit [2]. Error-prone process [2]. Side-tracked by short-term issues [1]. Less advance planning and estimation [2]. Inexperience [2]. |
| Constraints of Agile | Need good project management [1]. Team size should be smaller [1]. Complex [1]. Alignment with methodology [3]. May not be suited for entirely new product [1]. Requires constant communication [1]. Collaboration tools a must [1]. Agile expensive when project requires high- communication-mode (in definition or endgame) [1]. Partner readiness [1]. |

| Question | Responses [count of mentions] |
|---|---|
| Benefits of Agile | Managing complexity [2] and uncertainty [3]. Flexibility and responsiveness [4]. Quality [3]. Product-market fit [2]. Transparency [1]. Team participation [1]. |
| Projects/firms suited for Agile | Requirements ambiguity [3]. Complex applications [1]. Customer-facing products [2]. |
| Projects or firms best suited for Structured | Well-defined features [1] and timelines [1]. New products [1]. Traditional industries and public sector [1]. Need to interface with non-Agile teams [1]. Language or communication barriers [1]. Backend projects [2]. |

*Table 6: Interview Answers Summary*

All interviewees cited some form of arbitrage as a reason for outsourcing. Cost arbitrage was mentioned by all as a motivation in many questions; talent and the ability for vendors to flexibility scale up (and down) resources were other oft-cited arbitrage strategies. A few mentioned access to local knowledge for outsourcing nearshore and farshore but some others were unaware of this motivation.

We asked interviewees if they had seen positive outcomes from nearshoring and if they could explain such outcomes. All of them had observed positive results from nearshoring, in either some or all of the measures we tracked. One manager even had a direct comparable: when he began his development work in Paris, France, he had an outsourced nearshore team based in Eastern Europe. Subsequently, he moved his operations to Seattle, USA, thereby transforming the same development team to an farshore one. He observed decreased efficiency, increased *Project Management effort*, and more *communication problems*, despite the fact that he had worked with this same team for over a year and had established a strong rapport with them. Thus, it seems the added stress and overhead of temporal distance is an impediment to success. This sentiment was affirmed by other interviewees.

We asked interviewees if they had noticed increased costs arising from Agile development. Some had not and postulated that perhaps the question was improperly understood by the respondents. A plurality of interviewees, however, felt that increased costs (compared to budget) arising from Agile methodologies were plausible because Agile product specification tended to be less precise than Structured; as the Agile development unfolded, new features or other requirements would emerge, often as unbudgeted work. Worse, some work could be discarded. Many noted the increased cost of communication, often involving the entire team for meetings. Lack of experience with Agile was cited as another culprit. Overall, though changing requirements was considered a positive trait because it resulted in a final product that would meet end-user expectations.

Agile was lauded for its flexibility, transparency, product-market fit, suitability for complex projects, and team engagement; but it requires intensive communication and project management. The importance of training and investing in team alignment around methodologies and processes to mitigate ill effects was noted. Structured methods had their place for well-defined projects.

# 7   Discussion

Our research questions were:

RQ1: In the outsourced IT Software Engineering industry, what is the effect of outsourced development team location (nearshore vs. farshore) on: 1) *Meeting frequency;* 2) *Communication problems;*[6] 3) *Business day overlap;* 4) *Asynchronous communication;* 5) *Collaboration types;* 6) *Methodology;* 7) *Ways to define products;* 8) *Project management effort;* 9) *Success;* 10) *Cost;* 11) *Schedule;* 12) *Quality.*

RQ2: In the outsourced IT Software Engineering industry, what is the effect of software development methodology on: 1) *Location (nearshore vs. farshore);* 2) *Meeting frequency;* 3) *Communication problems;* 4) *Business day overlap;* 5) *Asynchronous communication;* 6) *Collaboration types;* 7) *Ways to define products;* 8) *Project management effort;* 9) *Success;* 10) *Cost;* 11) *Schedule;* 12) *Quality.*

RQ3: In the outsourced IT Software Engineering industry where the development methodology is Agile, what is the effect of outsourced development team location (nearshore vs. farshore) on: 1) *Meeting frequency;* 2) *Communication problems;* 3) *Business day overlap;* 4) *Asynchronous communication;* 5) *Collaboration*

---

[6] The boxed outcomes indicate statistically significant effects.

*types;* 6) *Ways to define products;* 7) *Project management effort;* 8) *Success;* 9) *Cost;* 10) *Schedule;* 11) *Quality.*

Our quantitative results indicate that outsourcing temporal distance is a greater determinant of outcomes than development methodology. Since *success* is correlated with other outcomes, temporal distance affects other outcome-related variables such as *cost*, lower *PM effort,* maintaining *schedule,* higher *quality,* and fewer *communication problems.* These findings are consistent with motivations for nearshore GSD, as noted in the literature and cited in section, *Outsourced Nearshore Software Development.* Development methodology, however, appears to influence higher *costs.* We thus confirm the relative insignificance of methodology on outsourced GSD.

In prior research of GSD, there has been a focus on outsourcing as a means for managers to arbitrage costs (both labor and infrastructure), obtain access to talent, and lower the risk of hiring expensive employees in advanced economies. To the extent that these advantages can be obtained either nearshore or farshore, outcomes can be meaningfully improved by preferring temporally-near locations—i.e., within five hours of the primary development center. Thus, lower costs, higher quality, and timely schedules can be better attained with nearshore outsourcing compared to farshore. This conclusion reflects similar findings by other researchers as described in the literature review.

Development methodology has a statistically significant effect on expected costs, in that budgets are more likely to be exceeded with an Agile methodology. As some interviewees pointed out, this might not be an adverse outcome, if the resulting project better meets the needs of the business. That said, some organizations may have strict controls on costs, in which case an Agile approach that could result in higher costs may be unattractive. To the extent that software development projects are well-suited to Structured methods—for instance, if requirements are static or are generally well-known—then an Agile approach could be declined with no loss because our findings indicate that methodology does not seem to affect outcomes.

While development methodology may not be a strong determinant of outcomes (some researchers found it was not an independent variable (Estler, et al., 2014)), a few highly desired attributes of Agile development, such as intensive communication and project management effort, may be undermined by choice of outsourced location. That is, farshore outsourcing seems to be associated with communication problems and less frequent meetings, both of which are quintessentially Agile. So, given management strategies such as farshore/nearshore or infrequent meetings/asynchronous communication, the development methodology may be determined.

Software development, as an intellectual activity with objective inputs and outputs, seems well-suited for outsourcing—yet temporal distances can affect outcomes. The implication of these findings is that nearshoring can increase the prospect of attaining positive outcomes; and, conversely, farshoring can decrease the likelihood. In any case, arbitrage strategies must be implemented with accommodations for those combinations of choices that are likely to cause problems. Particular care should be paid to alleviating likely problems in communication.

## 7.1 Further Research

In software development, it has long been posited that methodology is a determinant of success (Boehm, 1981); in particular, Agile methods for insourced, onsite development have been associated with successful outcomes (VersionOne, 2017). However, methodology does not appear to be a strong determinant of outcomes in the outsourced software development industry when the location is either farshore or nearshore. In studying determinants of outcomes, researchers should be aware that location can supersede the impact of methodology. This observation opens up a broader question: could location be a significant impact on insourced development, too? If not, to what extent does a firm's internal culture mitigate these potential impacts?

Distributed software development has been found to increase the cost of communication and risk of problems (Espinosa & Carmel, 2004). According to the Allen Curve, even small distances can impede the flow of communication in research and development (Allen, 1984). So, with distributed teams, does the deterioration in communication (and successful outcomes) increase monotonically from proximity to nearshore and thence to farshore as predicted by the Allen Curve, or does it perhaps fall off more precipitously with temporal distance? Allen has stated to the effect that quality of relationships determines the quantity of communication, not available technology (Allen & Henn, 2006). Is this true in software development, even with all the new social media-based tools available, such as Slack, JIRA, and others? And even though we have focused on temporal proximity, could geographic proximity further impact outcomes?

If methodology is not an independent variable or at least not a strong determinant of outcomes, it's possible that other variables—for example, type of communication—are. Other factors such as project size, project complexity, team experience could also be determinants. Research to determine independent variables that might affect methodologies and outcomes could be fruitful. Our research provides a starting point in identifying frequency of meeting, communication type, or ways to define products, as potential independent variables. Characterizing such impacts on methodology could help managers adapt methodologies to operational conditions.

## 7.2  Customer Implications

In outsourced software development, temporal distances matter. Our research indicates that nearshore is an adaptation to these distances that alleviates some deleterious effects. By the same token, it suggests that for farshore development, teams should not try to duplicate the communication practices that are successful when team members are nearby (e.g., onsite development).

For customers of outsourced offshore software development, the research indicates that they should select their outsourcing location carefully, favoring temporally nearer locations over farther ones, since time difference is a stronger determinant of success (and its related outcomes) than methodology. Furthermore, development methodology is not a highly significant determinant of success in outsourced software development—a project can be just as successful with traditional, structured management techniques. In fact, structured development seems to be more effective at keeping costs aligned to budgets. What this means in practice is that managers should have reasonable expectations about the broad applicability of Agile methods to outsourced farshore projects, since Agile requires intense synchronous communication. For nearshore development, we confirmed that Agile methodology has better *success*, lower *PM effort,* and higher *quality* at the cost of higher *costs* and *communication problems.* In Agile, the likelihood of *cost* increases was higher for near and far, but Agile farshore software development showed the largest increase.

Agile farshore development had lower *success*, greater *PM effort,* more *schedule* problems, and lower *quality*. On the other hand, farshore development may be successful given the proper accommodations, such as greater use of asynchronous communication and more formal definition of product requirements. In fact, one experienced manager claimed that some projects, perhaps those of a routine nature or with limited end-user engagement, may be best managed using Structured methodologies and outsourced farshore to optimize costs.

Finally, managers may wish to adapt their development methodology to accommodate management choices such as meeting frequency or intensive use of asynchronous communications. In other words, a strict application of Agile may not always be appropriate. *Table 7* summarizes the main findings for methodology and location of interest to customers of outsourced offshore software development.

| Software Development Methodology | |
|---|---|
| **Agile** | **Structured** |
| Higher costs (vs. budget) | More predictable costs (vs. budget) |
| **Outsourcing Location** | |

| Nearshore | Farshore |
|---|---|
| Higher success | More PM effort |
| Higher quality | Schedule risk |
| Lower schedule risk | Quality risk |
| Less PM effort | More communication problems |
| Fewer communication problems | |

*Table 7: Methodology and Location Findings*

## 7.3 Threats to Validity

Internal threats to validity include sampling methods, selection biases and survey construction. We discuss these in turn.

Sampling methods for finding respondents and interviewees may jeopardize validity. For example, among other sources, we used the connections on LinkedIn of the principal investigator, which could result in an unrepresentative sample. However, we partially addressed this by public postings on social media with relevant hashtags to reach different audiences.

Selection biases of respondents could also affect validity in that it's possible those with strongly held opinions would be more likely to respond to the survey or interview. Similarly, interviews may have been conducted with individuals not representative of customers. These threats are common to studies that employ the survey-and-interview methodology and often cannot be easily mitigated. However, our practice of asking qualifying questions to eliminate unqualified respondents is a step in that direction as was the heuristic of selecting interviewees from diverse backgrounds.

Survey *granularity* and respondent interpretation of both survey and interview questions are a hazard to validity. For example, we required respondents to select either *Nearshore* or *Farshore* as the outsourcing location, deliberately avoiding a hybrid in which an outsource team could be in more than one location. There was no option for finer gradations of *Nearshore,* such as a team located in the same time zone, but physically separated. Further, how respondents subjectively interpreted the survey's Likert items could affect validity: what is "low" to one person may well be "medium" to another. To reduce differences of interpretation, we provided examples or definitions of the Likert values. And, by forcing respondents to choose between *Nearshore* or *Farshore*, we intended that they would strive for internal consistency in answering subsequent questions. Our correlation tests suggest that ambiguity was minimal and interviewee answers consistency attained.

Among external threats to validity are survey *coverage* and *responsiveness*—that is, does the survey data permit generalization and did the non-respondents possess characteristics that were not present to the same degree in the respondents? We attempted to both survey and interview a diverse group of experts as described in *Selection of Participants,* so we believe risks to validity from coverage are minimized. We received 80 qualified responses from a broad range of roles and companies, which should mitigate the threat of poor responsiveness. Still, we cannot eradicate these risks.

*Interviewer effects* are a well-known threat to validity (Lavrakas, 2008). We attempted to minimize these by adhering to the interview question script. Since the interviews were conducted under a variety of conditions, such as in-person or via telephone, we were exacting about signaling when the formal interview began and ended, so that other discussion topics would not be commingled. We also declined to express opinions or provide leading commentary. However, for clarity's sake, we did periodically verbally summarize the interviewees' statements. Nevertheless, we cannot guarantee that interviewer effects were completely suppressed.

# 8   Conclusion

We studied the effects of temporal distance and methodology on outsourced GSD using a survey of 80 respondents, interviewed 6 of them, and analyzed the correlations between *location* or *methodology* and *success, cost, Project Management effort, schedule, quality, communication problems, team meeting, business day overlap, asynchronous communication, collaboration type,* and *ways to define products.* We found that *Nearshore* has a positive effect on *cost*, lower *Project*

*Management effort,* maintaining *schedule,* higher *quality,* and fewer *communication problems.* These findings are consistent with the literature on outsourcing nearshore (Oshri, et al., 2009). Šmite stated that "nearshoring is an alternative solution that may enable more successful collaboration, although it may not result in the cost-savings intended" (Šmite, et al., 2010). Our research, however, confirms the finding of improved communications in nearshore outsourced GSD and that there are still cost savings. In-depth interviews confirmed that customers experienced superior outcomes such as higher quality, greater flexibility, and efficiency with nearshore development.

We found that Agile development methodology influences higher *cost.* But we confirmed the outcomes in outsourced nearshore GSD that users of Agile have long assumed: greater success and quality compared to farshore. Since our study is limited to the outsourced GSD, its conclusions may not apply to insourced or onshore. Our methodology may have limitations such as interviewer bias, surveyee selection bias, survey granularity and others.

For customers, one useful insight is that if they are constrained to use an Agile methodology in outsourced software development, nearshoring is a better choice over farshoring.

In conclusion, we found advantages to nearshore development, especially in combination with Agile methods, but that costs could increase with Agile.

# 9   Appendix A: Survey Questions

Aside from a consent agreement, qualifying questions and contact questions, the survey consisted of the following (* indicates mandatory):

1. The number of employees in my company is:
    1 - 99
    100 - 999
    1000 - 9999
    10,000 or more

2. Which of the following best describes your IT job role?
    Software Developer
    Technical Lead or Architect
    Manager or Project Manager
    Product Owner
    CIO or similar executive level
    Other (please specify)

3. Our outsourced development team is primarily:
    Nearshore (i.e., 5 hours' time difference or less)
    Farshore (i.e., more than 5 hours' time difference)

4. We have development team meetings with outsourced team members:
    Daily or more frequently
    Three times per week
    Twice a week
    Once a week
    Less than once a week

5. Communication was a problem…
    | 1 – Almost never | 2 – Never to half | 3 – About half the time | 4 – Half to Most | 5 – Most of the time |
    | --- | --- | --- | --- | --- |

6. We see better results working with outsourced team members when:
    There is sufficient overlap in the business day to allow all team members to meet at least once
    There is enough overlap in the day to allow some team members to meet
    Team members don't need to meet daily, so overlaps aren't important.

7. Asynchronous communication (i.e., participants not always active at same time) using messaging or email with the outsourced team is:
    Sufficient for most of our development work (more than 80% of communication is done with these tools)
    Complementary for development work (40% - 80% of communication)
    Inadequate for most development (less than 40%)

8. Collaboration among insourced and outsourced team members is mostly:
    Asynchronous (i.e., participants not simultaneously active), using email or messaging
    Synchronous, using tools such as video conferencing
    More or less equally synchronous and asynchronous

9. Our software development methodology is:

Mostly Agile (i.e., iterative, where requirements & solutions evolve via collaboration between self-organizing cross-functional teams).
Mostly Structured software development (i.e., requirements and designs are defined in advance with formal interfaces and specifications).
Equally Agile and Structured

10. The most important ways we define product details and work to be done are (please rank):
    Shared documents (e.g., Word files or Wikis)
    Meetings (in person or virtual)
    Other (e.g., asynchronous)

11. The effort to manage outsourced projects is usually…
    | 1 – Less than expected | 2 – Somewhat less than expected | 3 – About expected | 4 – Somewhat more than expected | 5 – More than expected |

12. Our outsourced projects are usually successful (i.e., meet time, cost, feature objectives)…
    | 1 – Almost never | 2 – Never to half | 3 – About half the time | 4 – Half to Most | 5 – Most of the time |

13. The costs of our outsourced projects are higher than expected…
    | 1 – Almost never | 2 – Never to half | 3 – About half the time | 4 – Half to Most | 5 – Most of the time |

14. Our outsourced projects have schedule problems…
    | 1 – Almost never | 2 – Never to half | 3 – About half the time | 4 – Half to Most | 5 – Most of the time |

15. The quality of our outsourced projects is:
    1 – Low   2 – Low to Average   3 – Average   4 – Average to High   5 – High

# 10  Appendix B: Qualitative Interview Questions

## 10.1  Nearshore versus Farshore

1. What are the reasons for choosing outsourcing nearshore for your firm? For farshore?
    a. Is your firm motivated by cost, talent or other arbitrage strategies?
    b. Is your firm motivated by adaptation (to local market) strategies?
2. Do you use different locations for varying reasons?
    a. Is your firm organized to support outsourcing? Is it able to support farshore or nearshore equally well?
    b. Our research analysis suggests that outsourcing at nearshore locations results in better success, cost savings, schedules, higher quality and fewer communication problems.
    c. Have you experienced any of these results? Which ones?
    d. What are the reasons for them from the perspective of your experience or that of your firm?
3. What accommodations are used by your firm to outsource farshore? Nearshore?

## 10.2  Agile versus Structured

1. Agile is popular with many firms; 50 out of 80 stated they used it, with a further 17 claiming both it and structured. What are the reasons for your firm's chosen methodology?
2. According to our analysis, costs appear to be higher with Agile. What is your experience with this?
    a. Are there hidden costs?
    b. Are there constraints to the use of Agile? To the use of Agile with outsourcing?
    c. Are there benefits that are not accounted for?
3. Are there projects or companies that are better suited to use Agile than Structured?

# 11 Appendix C: Means and Standard Deviations

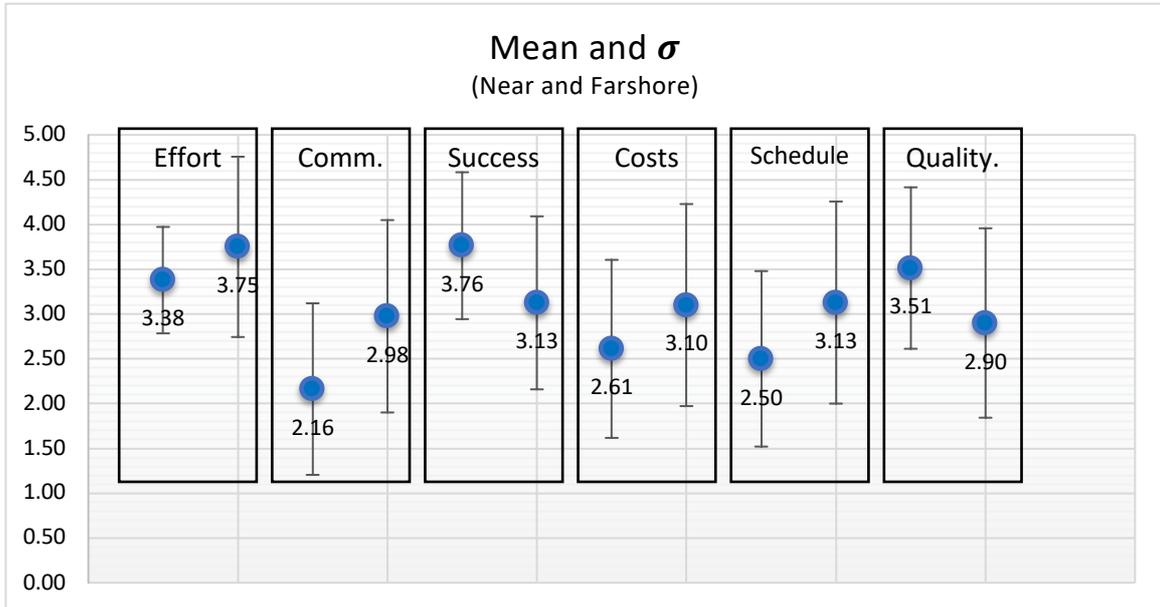

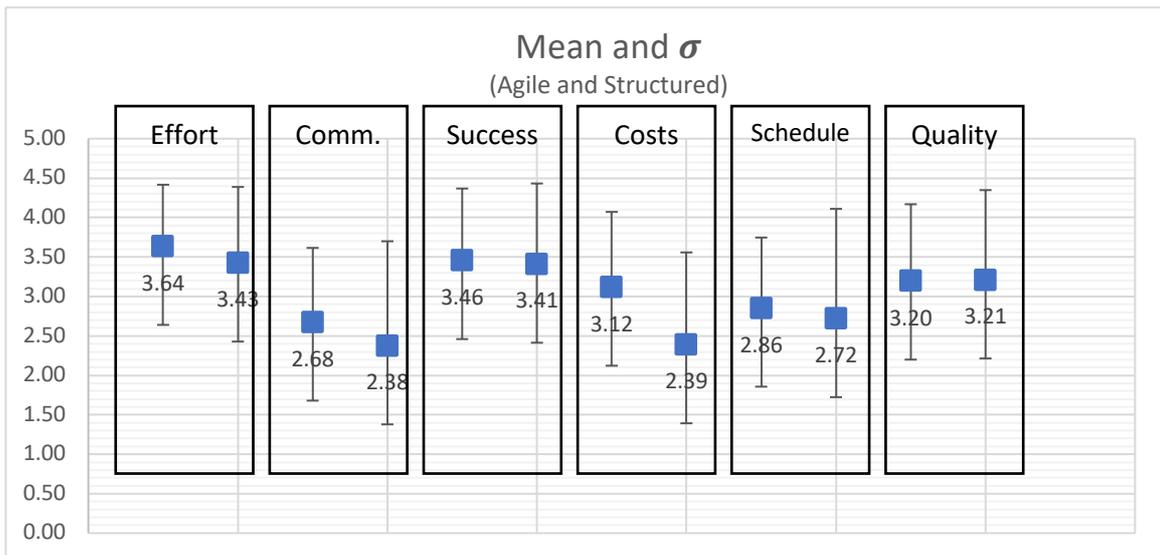

## 12 Conflict of Interest

On behalf of all authors, the corresponding author states that there is no conflict of interest.